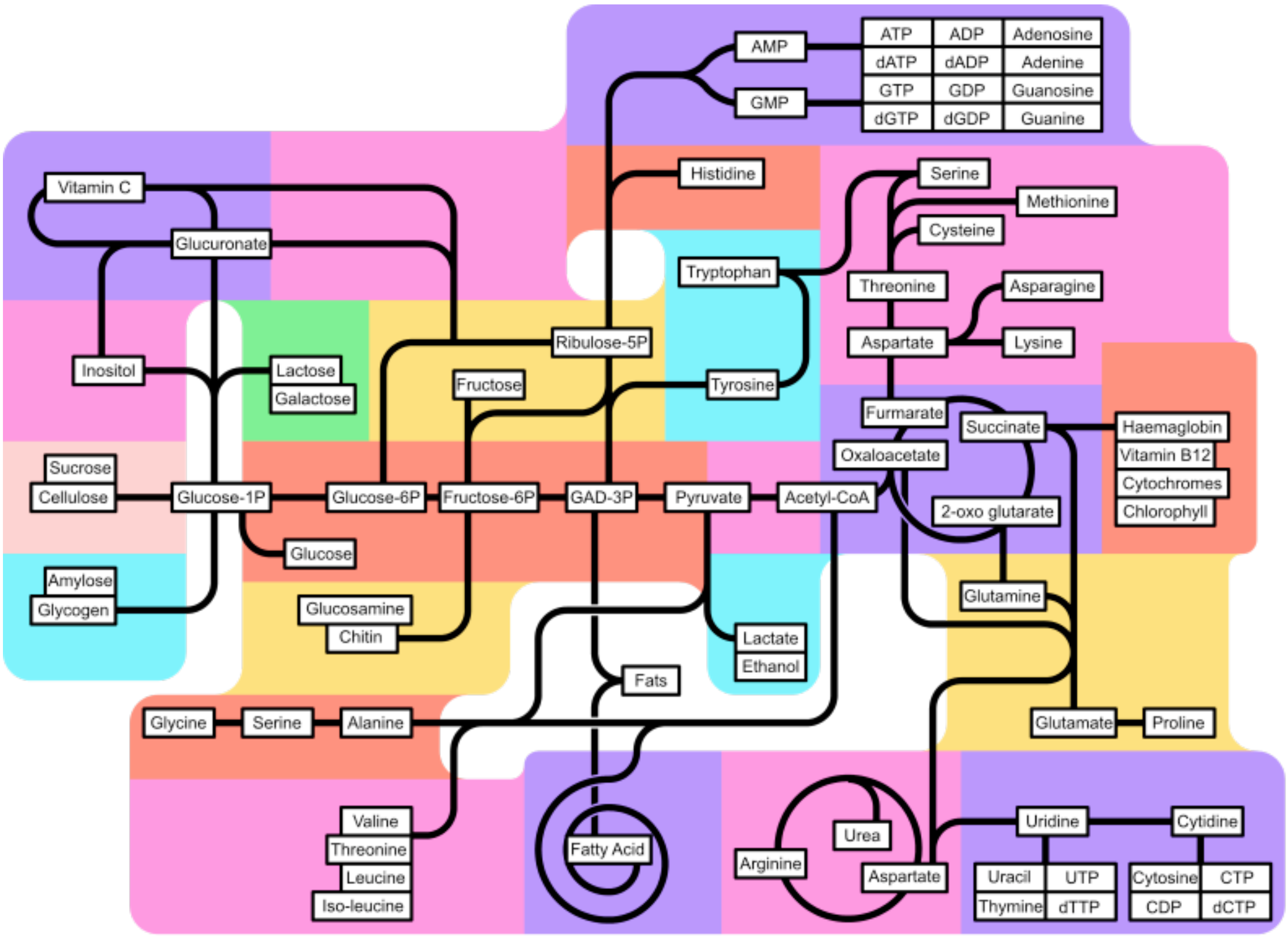

Andrej Poleev. Universal Metadata Standard.

Abstract. The basis of consciousness is an association of notions, the neuronal network. Similarly, the creation of a next generation internet (semantic web) is impossible without attributes, allowing the semantic association of documents and their integration into information context. To achieve these goals, the Universal Metadata Standard (ums) may be an ultimative tool, which could serve as a basis for documentography, and is functionally required for interpretation of documents by the automatic operating systems.

Key words: document, metadata, classification, identification, association, documentography, metagraphy, standard, metabase.

Универсальный стандарт метаданных. 8.03.2011

Резюме. Основой сознания является ассоциативная связь понятий, согласованная работа элементов нейрональной сети. Аналогично этому, создание интернета нового поколения (semantic web) невозможно без атрибутов, позволяющих осуществлять семантическую связь документов и интеграцию их в информационный контекст. Для реализации этих целей предлагается ввести Универсальный Стандарт Метаданных (universal metadata standard, ums), который мог бы служить основой документографии (documentography), функционально необходимой для интерпретации документов в автоматических операционных системах.

Ключевые слова: документ, метаданные, классификация, идентификация, ассоциация, документография, метаграфия, стандарт, база метаданных.

Цель научного познания – объять необъятное. Невозможность достижения этой цели очевидна, однако если принимать её за максиму научно–познавательной деятельности и путеводную звезду в поисках истины, удовлетворение стремления узнать болше и расширить индивидуальный круг знаний представляется вполне разумным и оправданным мотивом любого человека. В сущности, значительную часть времени люди заняты организацией информационного потока, непрерывно поступающего в их мозг через органы чувств и рецепторы как из тела так и извне. Не только их благополучие, но и шансы на выживание, определяются тем, насколько эффективно происходит такое упорядочивание информации, в результате чего сырьё нервных импульсов превращается в достоверное знание.

Появление и развитие сознания связано с совершенствованием средств коммуникации, основанием чего является знаковая передача информации, язык. Непрерывное совершенствование техники коммуникации, преодоление семантических барьеров методом проб и ошибок, привело к возникновению стандартов передачи и восприятия информации, примером чего является книгопечатание (полиграфия). Проделав значительный путь, полиграфическая техника привела к становлению микроэлектроники, которая не только улучшила качество и расширила область достоверного знания, но и ознаменовала собой возможность злонамеренного манипулирования сознанием, поскольку из поля зрения читателей и зрителей, т.е. реципиентов информации, стали исчезать сферы производства и распространения знания, а также те аспекты документов, которые недоступны непосредственному человеческому восприятию, однако могут или должны быть восприняты обрабатывающими информацию машинами (компъютерами). Восполнить возникший пробел восприятия представляется важной задачей информатики.

Рассмотрим пример того, как происходит организация знания. В центре внимания учёного сообщества находится проблема накопления, верификации и систематизации знаний, оформляемых в виде научных публикаций. Однако появлению всякой публикации предшествует значительная деятельность, как правило сокрытая от публики. Черновой вариант научных статей – лабораторный журнал – это не что иное как сборник протоколов о запланированных экспериментах и их результатах. Однако в идеальном случае он должен протоколировать всё, относящееся к проводимой научной работе и отражать всё, что происходит в лаборатории в хронологическом порядке, начиная от целеполагания, гипотезы, экспериментальной проверки, выводов, впечатления об увиденном и услышанном. В формальном плане, лабораторный журнал должен описывать документы различного формата: фотографии, тексты протоколов, тексты публикаций, видеозаписи конференций (lab meetings), указания на источники в интернете и т.д. Все эти документы должны быть связаны между собой, снабжены комментариями, и доступны для просмотра и каталогизации. Например, в хронологическом порядке друг за другом могут следовать эксперименты или мысли, относящиеся к различным темам: теоретическое иследование определённого вопроса и сбор соответствующей информации; написание статьи или книги на основании уже завершённой работы; планирование тематически разнородных экспериментов. В связи с этим эта тематическая разнородность должна быть отражена в списках тем, а также в возможности экстрагировать однородную (родственную) информацию посредством указателей (thematic tags) и ссылок (location tags).

Компьютер MacBook, которым я пользуюсь, является таким собранием разнородных документов, и предоставляет возможности их тематического объединения. Однако для их описания, воспроизведения или визуализации необходимо дополнительное программное обеспечение. File Maker только частично удовлетворяет потребности систематизации и описания: на данном этапе отсутствует приемлемая панель обозрения и возможность открывать и использовать документы внутри данной программы, не прибегая к дополнительным программам. Все эти дополнительные программы в идеальном случае должны быть встроены в качестве опций, а не разбросаны по разным местам: web editor, web browser, photoshop, file maker, pdf reader, video or photo visualiser, text editor и т.д..

В связи с обилием документальной основы научного сознания и познания, проблема документации и систематизации данных приобретает первостепенное значение. Обычно документы классифицируют по алфавиту, по дате, по теме, по проекту, по формату, по местонахождению (local folder, internet address). Для их идентификации служит дата, порядковый или систематический номер, имя (название). Например, изображения (images) имеют формат (file format) jpg, gif, png, psd; тексты (texts) имеют формат pdf, doc, txt. Формат документа – это его идентификационный признак (identification tag), необходимый для опознания в операционных системах и инициации программ (процессирования). Однако в каждом формате до сих пор отсутствует его систематическое описание, необходимое и достаточное для интеграции и переноса в другие описательные системы (например, при копировании из электронной библиотеки в персональный компьютер). Всякий документ отражает реальные предметы и события, является их описанием, отображает определённые качества. Однако фотография не сохраняет информации о размерах объекта, о его происхождении, истории, цели. Всё это в идеальном случае должно входитъ в метаинформационное дополнение документа, по крайней мере в виде ссылок. Однако увеличение количества документов и форматов не сопровождается совершенствованием технических возможностей их восприятия и систематизации. Вместо этого происходит размножение описательных систем (doi, ISBN, URN, PURL, ISNI и др.) и псевдонимов (aliasing). Так например, журнальная статья, как правило в форматах html или pdf, в описательной системе NCBI/NLM получает номер (PUBMED ID), добавляется резюме (abstract) с сопряжённым указанием на время публикации, название журнала, имён авторов, языка, ключевых слов. Необходимо однако, чтобы эта описательная метаинформация добавлялась непосредственно в документ в качестве дополнения или расширения, чтобы было возможно упорядочивание документа при перемещении его в другие описательные системы (например, при переводе на другой язык, или при использовании в другой базе данных), а история такого перемещения (например, при копировании из электронной библиотеки) отображалась бы в документе. Для достижения этой цели следует создать универсальный стандарт для всех типов документов, и договориться о том, какие опции будут присутствовать в каждом формате; как их будут заполнять или модифицировать; что не должно подвергаться изменению. Мне представляются очевидными нижеперечисленные опции метаинформационного описания документов:

имя * (preferably unique name)
формат (format)
дата создания (date)
классификационная система (classification system used)
идентификационный номер (identity number)
язык ** (language)
локализация или место происхождения (position, location)
источник или автор (creator, origin, source)

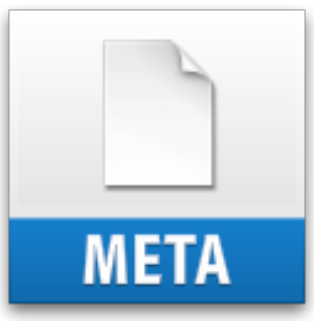

---

* Систематическое имя (systematic designation) – это последовательность символов (знаковая секвенция), на основании которой происходит идентификация обозначаемого объекта и установление соответствия между восприятием его посредством органов чувств (сенсорной репрезентацией) и языковой интерпретацией этого восприятия. Систематическое имя должно отличаться качествами, позволяющими отнести его к классу наименований, а также содержать необходимое дополнение, достаточное для однозначной идентификации среди родственных названий, обозначений и имён. Например, в узком круге лиц, достаточным является имя Андрей, в то время как в группе, имеющей в своём составе несколько людей с тем же именем, необходимо указывать родовое имя (фамилию) для раличения. В планетарном масштабе, достаточным является указание имени, даты и места рождения для установления идентичности. Систематическое имя для обозначения персон может состоять из двух или трёх имён, последовательности цифр, и географического определителя. Аналогично этому, систематическое имя организаций может содержать название, указание на дату и место основания, адрес, дату завершения деятельности. Ответ на 3 вопроса: Кто или что? Где? и Когда? является достаточным для идентификации и в других случаях.
Понятия каталог, номенклатура, классификация, регистр в значительной мере синонимичны, и обозначают список имён, объединённых в родственные группы, которые в свою очередь также сгруппированы на основании определённых критериев. Порядок группирования может изменяться в зависмости от выбранных критериев. Имена персон можно группировать по алфавиту, на основании даты или места рождения их прообразов. В динамичном пространстве категоризации, систематическое имя остаётся константой, кристаллизационным пунктом, отправной точкой в процессе семантической ассоциации, поиска и установления отношений и взаимосвязей между именами, понятиями, определениями, категориями.

** язык подразумевает знаковые системы естественных языков, имеющих дескриптивный и индикативный характер; языки программирования являются производными естественных языков, и имеют директивный характер алгоритмов, т.е. инструкций для автоматических операторов

Понятно, что инструкции по производству атомного оружия, или документы порнографического характера не могут быть доступны всем кому не лень. Поэтому для ограничения доступа к документам следует ввести градацию доступности.

Если документ будет претерпевать модификации (перенос в другую описательную систему, изменение размера, формата, названия), то первичные метаданные должны сохраняться, а изменения автоматически или мануально записываться: при переименовании добавляться синонимическое имя; в другой описательной системе (системе классификации) добавляться её обозначение и идентификационный номер в этой системе; при транспозиции записываться новый адрес в интернете или географическое соответствие, и т.д.

Для каждого атрибута УМС следует определить форму опции, дать её определение и формальное описание. Содержание каждой опции должно соответствовать правилам, на основании которых составлялся бы каталог допустимых значений (metabase: catalog of systematic designations). Например, авторство документов должно быть однозначным на основании списка авторов. Происхождение документа должно указываться на основании списка организаций. Указание типа документа (текст, рисунок, фотография, видео, звук), должно сопровождаться описанием (резюме), и типологической атрибутикой, характерной для каждого типа документов. Каждый документ должен содержать перечень объектов или явлений, отображением или описанием которых он является (биологический вид, астрономический объект, персона или группа лиц, организация, научная публикация и т.д.). Классификационная основа такого перечисления в настоящее время существует, (Encyclopedia of Life, International Plant Names Index, Catalogue of astronomical objects, PubMed, ICD и другие), следует использовать её в УМС.

Что же происходит в реальности? Рассмотрим показательный пример. Экстракция метаданных для документа octology.pdf, имеющего адрес http://www.enzymes.at/download/octology.pdf, дала следующий результат:

CreateDate = 2011:03:01 16:35:22Z
Title = octology
PageCount = 76
FileSize = 11 MB
Author = Max Madman
MIMEType = application/pdf
PDFVersion = 1.4
FileType = PDF
Creator = Pages
ModifyDate = 2011:03:01 16:35:22Z
PDFVersion (1) = 1.3
Producer = Mac OS X 10.5.2 Quartz PDFContext

Очевидна бессмысленность такого описания: указание на формат (pdf) присутствует 6 раз; кто создатель и автор документа – неясно; время создания и модификации документа совпадают и ничего не сообщают о времени его появления на свет божий. Пожалуй только указание на количество страниц и размер документа является осмысленным. Метаинформация включённых в текст иллюстраций (если таковая имелась) полностью утрачена в формате pdf. Публикация документа на портале Researchgate.net сопровождается указанием адреса DOI: details/Octology. Что это значит, неясно, поскольку проверка этого адреса в описательной системе DOI не приводит ни к какому результату. Хотя журнал, в котором осуществилась публикация данного документа, включён в базу данных NCBI/NLM, однако информация об этом документе до сих пор не появилась в электронной библиотеке PUBMED. Никакой возможности сделать это мануально нет: всё предоставлено программному обеспечению и автоматам, которые однако не в состоянии осуществить возложенную на них функцию. Сбой системы проявляется ещё и в том, что информация о документе с названием Octology, отсутствует в PUBMED/NCBI/NLM, но присутствует в связанной с нею описательной системе OCLC/WorldCat.

Ещё абсурднее выглядят метаданные документа, наугад выбранного из библиотеки PUBMED:
Palesch D, Sieńczyk M, Oleksyszyn J, Reich M, Wieczerzak E, Boehm BO, Burster T. Was the serine protease cathepsin G discovered by S. G. Hedin in 1903 in bovine spleen? Acta Biochim Pol. 2011 Mar 7, PMID: 21383996 (см. Приложение).

Подводя итоги, следует предложить программистам, терминологам, ISO и индустрии знаний разработать логически выверенную систему метаинформационного обеспечения для всеобщего пользования, чтобы производство документов не превратилось в самоцель, а приобрело бы надёжную основу, позволяющую на новом этапе социального и технического развития эффективно усваивать и организовывать знания. Параллельно, следовало бы дополнить существующие программы модулем, позволяющим визуализировать и редактировать метаданные, а также ввести в употребление универсальные программы для всех типов документов (metadata editors).

Более общей тематической идеей данной статьи является создание комплекса семантических стандартов, частью которых может стать УМС. Никола Тесла, заложивший технические основы для создания интернета на рубеже 19 и 20 веков, руководствовался мечтой об упразднении границ, препятствующих общению и познанию. Сегодня интернет, став реальностью, сам создаёт виртуальную реальность, на основании которой конструируется действительность, сознание и общество. Необходимо указать на связанные с этим опасности. Так например, смысловое содержание одного из центральных понятий интернета 3 поколения – онтология***, умысленно искажается в идеологических целях: Онтологиями в бизнесе стали обозначать логические схемы, разработанные для манипуляции сознания, вбивания заранее заданных стереотипов мышления, пропаганды групповых интересов. Написанные на непонятном широкой публике искусственном языке, онтологические схемы призваны осуществлять скрытый контроль над обществом узким кругом лиц, определяющих правила их написания. В связи с этим, семантический интернет может стать инструментом тоталитарного управления, имеющего глобальный характер. Понятно, что захват власти может осуществиться конспиративно, а сам тоталитарный процесс в таком случае будет вынесен за рамки юридического регулирования. Чтобы исключить злонамеренное использование технологии интернета, необходимо своевременно принять упреждающие меры. Предлагаемые в данной статье универсальные стандарты позволят избежатъ данного развития и сделать семантический интернет более осмысленным, реалистичным и доступным для регулирования широким кругом его пользователей.

---

*** Поскольку бытие (сущность) объектов проявляется в действии, то комплексное описание взаимодействия в данном множестве объектов даёт наиболее полное представление об изучаемой области. Онтологическая схема – это формализованое описание связей и взаимодействия между объектами в определённом множестве объектов. Примером применения онтологических схем может служить научная область, включающая всю совокупность изучаемых явлений и объектов, методов изучения и описания, гипотез и теорий. Другой пример: производственное предприятие, являющееся совокупностью оборудования (средств производства), технологического описания производства (методов производства), правил поведения персонала (инструкций управления предприятием) и других условий его функционирования.

В центре онтологической схемы находится описание объектов, включающее наименование или адресацию и установление свойственных им атрибутов (качеств и свойств проявления). Всякое описание основывается на систематизации, позволяющем отнести описываемый объект к группе объектов данной онтологической схемы. При этом атрибуты объектов могут приобретать более общий характер систематических категорий, на основании которых всё множество объектов распределяется на субгруппы. Например, во множестве предметов, некоторые из них могут быть шаровидной формы, отличаться по цвету и т.д. Таким образом, различение объектов происходит путём систематизации на основании индивидуальных признаков, а категоризация является рекурсивной операцией, выделяющей необходимые и достаточные признаки объектов, на основании которых осуществляется их систематизация и распределение внутри данного множества объектов.

Однако онтологические схемы могут не только описывать данность, но и активно влиять на объекты, определять их поведенческий модус посредством установления правил взаимодействия. Субъективный фактор онтологических схем наглядно проявляется в государственном управлении, основывающемся на неполном, искажённом, или неадекватном описании объектов, т.е. людей, социальных групп и их взаимоотношений, а также исключающем из рассмотрения онтологические схемы более общего характера (экология, биосфера, космология, философия). Неудивительно, что люди в таких онтологиях до сих пор рассматриваются как расходный материал, с которым можно обращаться как с неодушевлёнными предметами или домашним скотом.

Неосведомлённый читатель может восполнить пробел знаний, ознакомившись со следующими источниками:

Семантический интернет.
Michael K. Bergman. A Timeline of Information History.
Handbook of Metadata, Semantics and Ontologies, 2012, World Scientific Publishing.
Tom Heath and Christian Bizer (2011) Linked Data: Evolving the Web into a Global Data Space (1st edition). Synthesis Lectures on the Semantic Web: Theory and Technology, 1:1, 1-136. Morgan & Claypool. ISBN: 9781608454310.

Биоонтологии.
Онтология музыки.
Онтологии для электронного правительства.
Конструирование онтологий.

Метаданные документов.
Список форматов документов.
Adobe XMP.
Стандарты метаданных.
eXtended MetaData Registry (XMDR) Project.
Introduction to Metadata, 2008, by Tony Gill, Anne J. Gilliland, Maureen Whalen, and Mary S. Woodley, Murtha Baca (Ed.)
RIP Keywords Meta Tag.

Библиография о метаданных (метаграфия)
Steven J. Miller. Metadata and Cataloging Online Resources, 2010.
Greenberg, Jane. Metadata and Digital Information. In Marcia J. Bates, Mary Niles Maack, Miriam Drake eds. Encyclopedia of Library and Information Science, 2009, New York: Marcel Dekker, Inc.

Метаданные в искусстве, литературе и философии
Mark Amerika META/DATA A Digital Poetics, 2007, MIT Press.
Metaexhibition.
The Handbook of Organization Theory: Meta-theoretical Perspectives by Haridimos Tsoukas, Christian Knudsen. O.rd Un.ty Press, 2003, ISBN: 0199258325.
Metadata Symposium, Academy of Motion Picture Arts and Sciences.

Лабораторный журнал.
Amber Dance. How to choose your lab's next electronic lab notebook. The Scientist, 2010, vol. 24, 5, p. 71.

Технические средства организации научной литературы
EndNote.
Mekentosj Papers.

Приложение.

Экстракт метаданных упомянутого в тексте статьи документа из библиотеки PUBMED; экстракция произведена с помощью http://www.serversniff.net/file-info.php.

1. http://www.ncbi.nlm.nih.gov/pubmed/21383996

FileType(guessed) = XML
title - Was the serine protease cathepsin G discovered by ... [Acta Biochim Pol. 2011] - PubMed result
keywords - PubMed, National Center for Biotechnology Information, NCBI, United States National Library of Medicine, NLM, MEDLINE, Medical Journals, pub med, Entrez, Journal Articles, Citation search
description - PubMed is a service of the U.S. National Library of Medicine that includes over 19 million citations from MEDLINE and other life science journals for biomedical articles back to the 1950s. PubMed includes links to full text articles and other related resources.
author - pubmeddev
ncbi_stat = false
ncbi_phid = CE8875A4D78C969100000000001DA980
ncbi_pdid = abstract
Keywords = PubMed, National Center for Biotechnology Information, NCBI, United States National Library of Medicine, NLM, MEDLINE, Medical Journals, pub med, Entrez, Journal Articles, Citation search
Title = Was the serine protease cathepsin G discovered by ... [Acta Biochim Pol. 2011] - PubMed result
ncbi_pagesize = 20
ncbi_filter = all
FileSize = 92 kB
ncbi_format = html
ncbi_uidlist = 21383996
Author = pubmeddev
ncbi_report = abstract
ncbi_resultcount = 1
MIMEType = text/html
ncbi_app = entrez
ncbi_hitstat = true
FileType = HTML
ncbi_pageno = 1
ncbi_db = pubmed
ncbi_sortorder = default
ncbi_sessionid = CE8875A4D78C9AD1_0029SID
Robots = index,nofollow,noarchive
ncbi_op = retrieve
Description = PubMed is a service of the U.S. National Library of Medicine that includes over 19 million citations from MEDLINE and other life science journals for biomedical articles back to the 1950s. PubMed includes links to full text articles and other related resources.

2. http://www.actabp.pl/pdf/Preprint/20112013.pdf

FileType(guessed) = PDF document, version 1.4

format - PDF 1.4

mimetype - application/pdf

MetadataDate = 2011:03:06 19:04:38+01:00

ManifestReferenceDocumentID (7) = xmp.did:E981B13A4B44E0118B25D8B18DDAEA80

ManifestReferenceInstanceID (6) = xmp.iid:E881B13A4B44E0118B25D8B18DDAEA80

ManifestPlacedResolutionUnit (8) = Inches

ManifestLinkForm (4) = ReferenceStream

DocumentID = xmp.did:36A96DDD4444E011BA44C4547900889D

ManifestLinkForm (7) = ReferenceStream

ThumbnailHeight = 256

ManifestLinkForm (9) = ReferenceStream

ManifestReferenceDocumentID (3) = xmp.did:ED81B13A4B44E0118B25D8B18DDAEA80

ManifestPlacedResolutionUnit = Inches

MIMEType = application/pdf

FileType = PDF

ManifestReferenceDocumentID (5) = xmp.did:2F33EFA24A44E0118B25D8B18DDAEA80

ManifestPlacedYResolution (8) = 300.00

ManifestReferenceInstanceID (5) = xmp.iid:3333EFA24A44E0118B25D8B18DDAEA80

ManifestPlacedResolutionUnit (3) = Inches

DerivedFromOriginalDocumentID = adobe:docid:indd:fac21e1b-e9a1-11db-bb63-f87ecfa109dc

ManifestPlacedResolutionUnit (5) = Inches

ManifestPlacedXResolution (7) = 300.00

ManifestLinkForm = ReferenceStream

ManifestReferenceDocumentID (8) = xmp.did:EB81B13A4B44E0118B25D8B18DDAEA80

HistoryWhen = 2010:01:08 10:37:37+01:00, 2010:01:08 10:37:37+01:00, 2010:01:08 10:38:47+01:00, 2010:01:09 20:37:04+01:00, 2010:02:15 17:38:17+01:00, 2010:02:15 17:38:18+01:00, 2010:02:15 17:54:28+01:00, 2010:02:15 18:09:18+01:00, 2010:02:15 18:10:28+01:00, 2010:02:15 18:28:53+01:00, 2010:02:15 18:35:15+01:00, 2010:02:15 18:54:59+01:00, 2010:02:15 18:57:27+01:00, 2010:02:15 23:31:26+01:00, 2010:02:15 23:34:43+01:00, 2010:02:15 23:43:09+01:00, 2010:02:15 23:47:48+01:00, 2010:02:15 23:51:01+01:00, 2010:02:15 23:52:57+01:00, 2010:02:15 23:58:11+01:00, 2010:02:16 00:00:58+01:00, 2010:02:16 00:03:51+01:00, 2010:02:16 10:16:48+01:00, 2010:02:16 10:26:27+01:00, 2010:02:16 11:03:51+01:00, 2010:02:16 11:28:37+01:00, 2010:02:16 15:07:05+01:00, 2010:02:16 15:09:02+01:00, 2010:02:16 15:13:01+01:00, 2010:02:16 15:21:18+01:00, 2010:02:16 15:35:34+01:00, 2010:02:16 15:44:47+01:00, 2010:02:16 18:18:01+01:00, 2010:02:16 18:58:14+01:00, 2010:02:16 18:58:36+01:00, 2010:02:16 19:19:28+01:00, 2010:02:16 19:48:35+01:00, 2010:02:16 19:57:42+01:00, 2010:02:16 20:25:24+01:00, 2010:02:16 22:38:42+01:00, 2010:02:24 12:57:22+01:00, 2010:02:24 12:57:22+01:00, 2010:02:24 12:59:57+01:00, 2010:02:24 13:06:53+01:00, 2010:02:24 16:25:01+01:00, 2010:02:24 16:27:04+01:00, 2010:02:24 19:56:39+01:00, 2010:02:24 19:56:50+01:00, 2010:02:24 19:56:50+01:00, 2010:02:28 19:45:50+01:00, 2010:02:28 19:45:50+01:00, 2010:02:28 19:46:03+01:00, 2010:02:28 19:46:03+01:00, 2010:02:28 20:56:51+01:00, 2010:02:28 20:57:03+01:00, 2010:02:28 20:57:03+01:00, 2010:02:28 20:58:25+01:00, 2010:03:08 10:02:08+01:00, 2010:03:08 10:02:08+01:00, 2010:03:09 21:12:26+01:00, 2010:03:09 21:12:26+01:00, 2010:03:13 00:20:22+01:00, 2010:03:13

00:20:22+01:00, 2010:03:16 13:56:03+01:00, 2010:03:16 13:56:03+01:00, 2010:05:30 20:56:20+02:00, 2010:05:30 20:56:20+02:00, 2010:06:06 22:20:49+02:00, 2010:06:06 22:20:49+02:00, 2010:06:06 22:24:35+02:00, 2010:06:06 22:24:35+02:00, 2010:08:19 00:08:54+02:00, 2010:08:19 00:08:54+02:00, 2010:12:09 17:48:01+01:00, 2010:12:09 17:48:01+01:00, 2010:12:09 17:55:34+01:00, 2010:12:16 19:29:56+01:00, 2010:12:16 19:29:56+01:00, 2010:12:16 19:30:28+01:00, 2010:12:16 19:30:28+01:00, 2011:02:28 18:04:14+01:00, 2011:02:28 18:04:14+01:00, 2011:03:01 21:45:38+01:00, 2011:03:01 21:45:38+01:00, 2011:03:01 21:50:09+01:00, 2011:03:01 21:50:09+01:00, 2011:03:01 21:51:10+01:00, 2011:03:01 21:53:16+01:00, 2011:03:01 22:00:52+01:00, 2011:03:01 22:16:38+01:00, 2011:03:01 22:23:24+01:00, 2011:03:01 22:47:01+01:00, 2011:03:01 22:55:42+01:00, 2011:03:01 22:57:12+01:00, 2011:03:06 19:03:55+01:00

ManifestPlacedYResolution (3) = 300.00

RenditionClass = proof:pdf

DerivedFromRenditionClass = default

ManifestPlacedXResolution (3) = 300.00

Trapped = False

Trapped (1) = False

ManifestLinkForm (8) = ReferenceStream

ManifestReferenceInstanceID = xmp.iid:3733EFA24A44E0118B25D8B18DDAEA80

ManifestPlacedXResolution (1) = 150.00

ManifestPlacedResolutionUnit (7) = Inches

Producer (1) = Adobe PDF Library 9.0

ManifestPlacedYResolution (5) = 300.00

ManifestReferenceInstanceID (7) = xmp.iid:EA81B13A4B44E0118B25D8B18DDAEA80

ManifestLinkForm (3) = ReferenceStream

DerivedFromDocumentID = xmp.did:B3F060712F09E011A376BF5163C1EA45

OriginalDocumentID = adobe:docid:indd:fac21e1b-e9a1-11db-bb63-f87ecfa109dc

PDFVersion = 1.4

ManifestReferenceDocumentID (6) = xmp.did:E781B13A4B44E0118B25D8B18DDAEA80

ManifestPlacedYResolution (2) = 150.00

ManifestReferenceDocumentID (2) = xmp.did:D1A28B0814ABDF11BE44F632C2D0BD91

ManifestPlacedYResolution (9) = 300.00

ManifestReferenceInstanceID (3) = xmp.iid:EE81B13A4B44E0118B25D8B18DDAEA80

ManifestPlacedXResolution (5) = 300.00

ManifestLinkForm (1) = ReferenceStream

ManifestPlacedXResolution (8) = 300.00

DerivedFromInstanceID = xmp.iid:35A96DDD4444E011BA44C4547900889D

ModifyDate = 2011:03:06 19:04:38+01:00

ManifestPlacedResolutionUnit (2) = Inches

ManifestPlacedResolutionUnit (6) = Inches

HistoryAction = saved, saved, saved, saved, saved, saved, saved, saved, saved, saved, saved, saved, saved, saved, saved, saved, saved, saved, saved, saved, saved, saved, saved, saved, saved, saved, saved, saved, saved, saved, saved, saved, saved, saved, saved, saved, saved, saved, saved, saved, saved, saved, saved, saved, saved, saved, saved, saved, saved, saved, saved, saved, saved, saved, saved, saved, saved, saved, saved, saved, saved, saved, saved, saved, saved, saved, saved, saved, saved, saved, saved, saved, saved, saved, saved, saved, saved, saved, saved, saved, saved, saved, saved, saved, saved, saved, saved, saved, saved, saved, saved, saved, saved, saved, saved, saved, saved, saved

ManifestPlacedXResolution (9) = 300.00

XMPToolkit = Adobe XMP Core 4.2.2-c063 53.352624, 2008/07/30-18:12:18

CreateDate (1) = 2011:03:06 19:04:22+01:00

XMP = SCALAR(0xa08aba0)

ThumbnailImage = SCALAR(0xa673a78)

ManifestReferenceDocumentID (4) = xmp.did:E581B13A4B44E0118B25D8B18DDAEA80

ManifestPlacedResolutionUnit (9) = Inches

Format = application/pdf

ManifestReferenceDocumentID (1) = xmp.did:D1A28B0814ABDF11BE44F632C2D0BD91

ManifestPlacedYResolution (7) = 300.00

PageCount = 6

ManifestPlacedYResolution (6) = 300.00

HistorySoftwareAgent = Adobe InDesign 6.0, Adobe InDesign 6.0, Adobe InDesign 6.0, Adobe InDesign 6.0, Adobe InDesign 6.0, Adobe InDesign 6.0, Adobe InDesign 6.0, Adobe InDesign 6.0, Adobe InDesign 6.0, Adobe InDesign 6.0, Adobe InDesign 6.0, Adobe InDesign 6.0, Adobe InDesign 6.0, Adobe InDesign 6.0, Adobe InDesign 6.0, Adobe InDesign 6.0, Adobe InDesign 6.0, Adobe InDesign 6.0, Adobe InDesign 6.0, Adobe InDesign 6.0, Adobe InDesign 6.0, Adobe InDesign 6.0, Adobe InDesign 6.0, Adobe InDesign 6.0, Adobe InDesign 6.0, Adobe InDesign 6.0, Adobe InDesign 6.0, Adobe InDesign 6.0, Adobe InDesign 6.0, Adobe InDesign 6.0, Adobe InDesign 6.0, Adobe InDesign 6.0, Adobe InDesign 6.0, Adobe InDesign 6.0, Adobe InDesign 6.0, Adobe InDesign 6.0, Adobe InDesign 6.0, Adobe InDesign 6.0, Adobe InDesign 6.0, Adobe InDesign 6.0, Adobe InDesign 6.0, Adobe InDesign 6.0, Adobe InDesign 6.0, Adobe InDesign 6.0, Adobe InDesign 6.0, Adobe InDesign 6.0, Adobe InDesign 6.0, Adobe InDesign 6.0, Adobe InDesign 6.0, Adobe InDesign 6.0, Adobe InDesign 6.0, Adobe InDesign 6.0, Adobe InDesign 6.0, Adobe InDesign 6.0, Adobe InDesign 6.0, Adobe InDesign 6.0, Adobe InDesign 6.0, Adobe InDesign 6.0, Adobe InDesign 6.0, Adobe InDesign 6.0, Adobe InDesign 6.0, Adobe InDesign 6.0, Adobe InDesign 6.0, Adobe InDesign 6.0, Adobe InDesign 6.0, Adobe InDesign 6.0, Adobe InDesign 6.0, Adobe InDesign 6.0, Adobe InDesign 6.0, Adobe InDesign 6.0, Adobe InDesign 6.0, Adobe InDesign 6.0, Adobe InDesign 6.0, Adobe InDesign 6.0, Adobe InDesign 6.0, Adobe InDesign 6.0, Adobe InDesign 6.0, Adobe InDesign 6.0, Adobe InDesign 6.0, Adobe InDesign 6.0, Adobe InDesign 6.0, Adobe InDesign 6.0, Adobe InDesign 6.0, Adobe InDesign 6.0, Adobe InDesign 6.0, Adobe InDesign 6.0, Adobe InDesign 6.0, Adobe InDesign 6.0, Adobe InDesign 6.0, Adobe InDesign 6.0, Adobe InDesign 6.0, Adobe InDesign 6.0, Adobe InDesign 6.0, Adobe InDesign 6.0, Adobe InDesign 6.0

DocChangeCount = 1551

ManifestPlacedXResolution (4) = 300.00

ManifestPlacedXResolution (6) = 300.00

Creator = Adobe InDesign CS4 (6.0.6)

ManifestReferenceInstanceID (1) = xmp.iid:D2A28B0814ABDF11BE44F632C2D0BD91

ManifestPlacedYResolution (1) = 150.00

ManifestPlacedResolutionUnit (4) = Inches

Producer = Adobe PDF Library 9.0

ManifestReferenceInstanceID (8) = xmp.iid:EC81B13A4B44E0118B25D8B18DDAEA80

ManifestReferenceDocumentID (9) = xmp.did:3433EFA24A44E0118B25D8B18DDAEA80

CreateDate = 2011:03:06 19:04:22+01:00
ManifestLinkForm (5) = ReferenceStream
ManifestLinkForm (6) = ReferenceStream
ManifestPlacedXResolution (2) = 150.00
InstanceID = uuid:4d5fb0f5-a423-4f60-8bac-78e8496af6c1
ManifestPlacedYResolution (4) = 300.00
ManifestReferenceInstanceID (2) = xmp.iid:D2A28B0814ABDF11BE44F632C2D0BD91
ManifestPlacedResolutionUnit (1) = Inches
FileSize = 1418 kB
ModifyDate (1) = 2011:03:06 19:04:38+01:00
ManifestLinkForm (2) = ReferenceStream
ManifestReferenceInstanceID (9) = xmp.iid:3533EFA24A44E0118B25D8B18DDAEA80
ManifestReferenceDocumentID = xmp.did:3633EFA24A44E0118B25D8B18DDAEA80
HistoryChanged = /, /metadata, /, /, /metadata, /;/metadata, /, /, /, /, /, /, /, /, /, /, /, /, /, /, /, /, /, /, /, /, /, /, /, /, /, /, /, /, /, /, /, /, /, /, /, /metadata, /;/metadata, /, /, /, /, /, /metadata, /;/metadata, /, /metadata, /metadata, /;/metadata, /, /metadata, /;/metadata, /, /, /metadata, /, /metadata, /, /metadata, /, /metadata, /, /metadata, /, /metadata, /, /metadata, /, /metadata, /, /metadata, /, /, /metadata, /metadata, /;/metadata, /, /metadata, /, /metadata, /metadata, /;/metadata, /, /, /, /, /, /, /, /
ManifestPlacedXResolution = 300.00
ManifestReferenceInstanceID (4) = xmp.iid:E681B13A4B44E0118B25D8B18DDAEA80
ThumbnailFormat = JPEG
ThumbnailWidth = 256
ManifestPlacedYResolution = 300.00
HistoryInstanceID = xmp.iid:8186C37439FCDE11A688BAD078C173F1, xmp.iid:8286C37439FCDE11A688BAD078C173F1, xmp.iid:8386C37439FCDE11A688BAD078C173F1, xmp.iid:A34CAC5D56FDDE11939ABE5EF03E0266, xmp.iid:61442185501ADF11A593DD4FF33579F7, xmp.iid:62442185501ADF11A593DD4FF33579F7, xmp.iid:63442185501ADF11A593DD4FF33579F7, xmp.iid:780157DA541ADF118E92E3EA29F7DFB6, xmp.iid:790157DA541ADF118E92E3EA29F7DFB6, xmp.iid:7A0157DA541ADF118E92E3EA29F7DFB6, xmp.iid:D8FD8D7A581ADF11870698B18DB54D83, xmp.iid:D9FD8D7A581ADF11870698B18DB54D83, xmp.iid:DAFD8D7A581ADF11870698B18DB54D83, xmp.iid:F0B87C3D6A1ADF118328FFEBB453B0C9, xmp.iid:58C5FB4F821ADF118328FFEBB453B0C9, xmp.iid:59C5FB4F821ADF118328FFEBB453B0C9, xmp.iid:5AC5FB4F821ADF118328FFEBB453B0C9, xmp.iid:5BC5FB4F821ADF118328FFEBB453B0C9, xmp.iid:5CC5FB4F821ADF118328FFEBB453B0C9, xmp.iid:5DC5FB4F821ADF118328FFEBB453B0C9, xmp.iid:5EC5FB4F821ADF118328FFEBB453B0C9, xmp.iid:5FC5FB4F821ADF118328FFEBB453B0C9, xmp.iid:60C5FB4F821ADF118328FFEBB453B0C9, xmp.iid:61C5FB4F821ADF118328FFEBB453B0C9, xmp.iid:62C5FB4F821ADF118328FFEBB453B0C9, xmp.iid:98C5550BE61ADF118328FFEBB453B0C9, xmp.iid:53142C90041BDF11AB57D4FD8D6B129F, xmp.iid:54142C90041BDF11AB57D4FD8D6B129F, xmp.iid:55142C90041BDF11AB57D4FD8D6B129F, xmp.iid:56142C90041BDF11AB57D4FD8D6B129F, xmp.iid:57142C90041BDF11AB57D4FD8D6B129F, xmp.iid:58142C90041BDF11AB57D4FD8D6B129F, xmp.iid:59142C90041BDF11AB57D4FD8D6B129F, xmp.iid:5B142C90041BDF11AB57D4FD8D6B129F, xmp.iid:5C142C90041BDF11AB57D4FD8D6B129F, xmp.iid:86461FD2271BDF11AB57D4FD8D6B129F, xmp.iid:87461FD2271BDF11AB57D4FD8D6B129F, xmp.iid:88461FD2271BDF11AB57D4FD8D6B129F, xmp.iid:8B461FD2271BDF11AB57D4FD8D6B129F, xmp.iid:C5FF36A7431BDF11BBC6DCC972B6B7DF, xmp.iid:575E8AC43B21DF118E14AB5AA9CB3F6F,

xmp.iid:585E8AC43B21DF118E14AB5AA9CB3F6F, xmp.iid:595E8AC43B21DF118E14AB5AA9CB3F6F, xmp.iid:5A5E8AC43B21DF118E14AB5AA9CB3F6F, xmp.iid:5B5E8AC43B21DF118E14AB5AA9CB3F6F, xmp.iid:5C5E8AC43B21DF118E14AB5AA9CB3F6F, xmp.iid:364D8BC67521DF118E14AB5AA9CB3F6F, xmp.iid:374D8BC67521DF118E14AB5AA9CB3F6F, xmp.iid:384D8BC67521DF118E14AB5AA9CB3F6F, xmp.iid:1385E8E69824DF119D46DE32DF47E67A, xmp.iid:1485E8E69824DF119D46DE32DF47E67A, xmp.iid:1585E8E69824DF119D46DE32DF47E67A, xmp.iid:1685E8E69824DF119D46DE32DF47E67A, xmp.iid:846996C39F24DF119D46DE32DF47E67A, xmp.iid:856996C39F24DF119D46DE32DF47E67A, xmp.iid:866996C39F24DF119D46DE32DF47E67A, xmp.iid:876996C39F24DF119D46DE32DF47E67A, xmp.iid:D3719246912ADF11AFA6E6294AFB0521, xmp.iid:D4719246912ADF11AFA6E6294AFB0521, xmp.iid:F1AE3C17572BDF1197D5DCFA11AF9412, xmp.iid:F2AE3C17572BDF1197D5DCFA11AF9412, xmp.iid:2068DB752C2EDF11A84DC5F88F067B6A, xmp.iid:2168DB752C2EDF11A84DC5F88F067B6A, xmp.iid:593550E0F730DF11B0B9918CD7D44B50, xmp.iid:5A3550E0F730DF11B0B9918CD7D44B50, xmp.iid:E59510561A6CDF11A4818CCF7EDFE615, xmp.iid:E69510561A6CDF11A4818CCF7EDFE615, xmp.iid:12250FFFA871DF118A6AE0FCC3D6E0B3, xmp.iid:13250FFFA871DF118A6AE0FCC3D6E0B3, xmp.iid:14250FFFA871DF118A6AE0FCC3D6E0B3, xmp.iid:15250FFFA871DF118A6AE0FCC3D6E0B3, xmp.iid:4601A8BE14ABDF1183A2BE947806EC8B, xmp.iid:4701A8BE14ABDF1183A2BE947806EC8B, xmp.iid:82ADCB15B403E0118920DC76910A31A0, xmp.iid:83ADCB15B403E0118920DC76910A31A0, xmp.iid:84ADCB15B403E0118920DC76910A31A0, xmp.iid:B0F060712F09E011A376BF5163C1EA45, xmp.iid:B1F060712F09E011A376BF5163C1EA45, xmp.iid:B2F060712F09E011A376BF5163C1EA45, xmp.iid:B3F060712F09E011A376BF5163C1EA45, xmp.iid:A3975BC55C43E0119CDDC27F8C0DC1DB, xmp.iid:A4975BC55C43E0119CDDC27F8C0DC1DB, xmp.iid:33A96DDD4444E011BA44C4547900889D, xmp.iid:34A96DDD4444E011BA44C4547900889D, xmp.iid:35A96DDD4444E011BA44C4547900889D, xmp.iid:36A96DDD4444E011BA44C4547900889D, xmp.iid:37A96DDD4444E011BA44C4547900889D, xmp.iid:38A96DDD4444E011BA44C4547900889D, xmp.iid:39A96DDD4444E011BA44C4547900889D, xmp.iid:3AA96DDD4444E011BA44C4547900889D, xmp.iid:3BA96DDD4444E011BA44C4547900889D, xmp.iid:3CA96DDD4444E011BA44C4547900889D, xmp.iid:76B300A74E44E011BA44C4547900889D, xmp.iid:77B300A74E44E011BA44C4547900889D, xmp.iid:FD6F79CF1648E011BFA6AC29DFBD2A96
CreatorTool = Adobe InDesign CS4 (6.0.6)

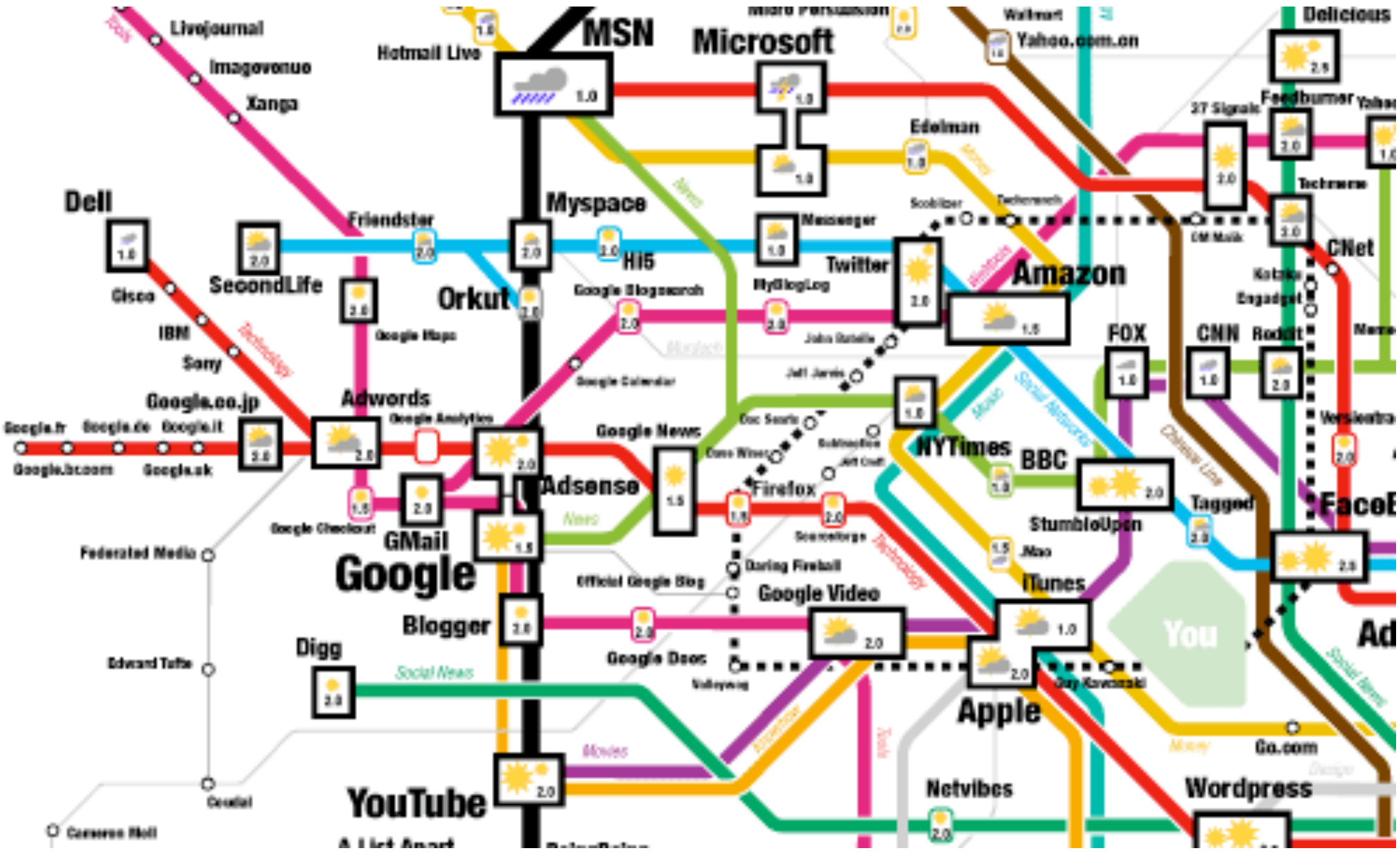